\documentclass[fleqn,twoside]{article}
\usepackage{espcrc2}
\newcommand{\eq}{\begin{eqnarray}}
\newcommand{\en}{\end{eqnarray}}

\newcommand{\bdm}{\begin{displaymath}}
\newcommand{\edm}{\end{displaymath}}

\newcommand{\be}{\begin{equation}}
\newcommand{\ee}{\end{equation}}
\newcommand{\bea}{\begin{eqnarray}}
\newcommand{\eea}{\end{eqnarray}}
\newcommand{\fs}{\; \; .}
\newcommand{\co}{\; \; ,}

\newcommand{\scs}{\co \;}
\newcommand{\sem}{ \; \; ; \;\;}

\newcommand{\dm}{\mbox{${\delta m^2}$}}
\newcommand{\dg}{\mbox{$\delta g$}}
\newcommand{\mueff}{\mu_{\mbox{\tiny{eff}}}}

\title{Hadronic processes and electromagnetic corrections
}
\author{I. Scimemi,
\address{ECM, University of Barcelona, Diagonal 647, E-08028
  Barcelona, Spain}}

\begin{document}
\begin{abstract}
The inclusion of electromagnetism in a low energy effective theory
is worth further study in view of the present  high precision experiments (muon $g-2$, $\pi_0\rightarrow \gamma \gamma$, $\tau$ decays, etc.). 
In particular in many applications of chiral perturbation theory, one 
has to purify physical matrix elements from electromagnetic effects. 
The theoretical problems that I want to point out  here are following:
the splitting of a pure QCD and a pure electromagnetic part in a  hadronic
process is model dependent: is it possible to parametrise in a clear way
this splitting?  What kind of information (scale dependence, gauge
dependence,..) is actually included in the parameters of the low
energy effective theory?
I will  attempt to answer these questions introducing  a  possible convention
to perform the splitting between strong and electromagnetic parts in
some examples.
\\
Talk given at SIGHAD2003, Pisa (Italy) October 8-10,2003.
\end{abstract}
\maketitle

\section{Introduction}
\label{sec:intro} 
The low energy effective theory of the Standard Model in the hadron sector  is the Chiral Perturbation Theory (ChPT). The chiral Lagrangian has been enlarged
 in order to include also electromagnetic effects in the meson sector \cite{Urech,Neufeld} and then also baryons\cite{Meissner} and leptons  \cite{Knecht:lep}--\cite{Cirigliano:lep2}.

The effective Lagrangian with virtual photons has been used
 to study
 isospin breaking corrections in the meson  and
 baryon
 sectors (see, e.g., Refs. \cite{Meissner,MMS,KU}), including 
hadronic atoms
  \cite{Bern}. 
Other applications are  the
evaluation of  isospin-breaking
 corrections in radiative $\tau$ decays, which is relevant 
for the analysis of
 the anomalous magnetic moment of the muon \cite{Cirigliano:radtau}, and the
 construction of  the  chiral Lagrangian 
 in the intrinsic parity odd sector at $O(e^2p^4)$, see 
Ref. \cite{Ananthanarayan}.
In this last reference, electromagnetic corrections to $\pi^0\rightarrow
\gamma\gamma$ were evaluated as well (see also \cite{GBH:02}).

In order to illustrate the object of the present work let us consider the example of the decay of  $\eta\rightarrow 3\pi$ in the framework of QCD \cite{eta3pi}. The amplitude for this
decay is proportional to $1/Q^2$, where 
\bea
Q^2=\frac{m_s^2-\hat{m}^2}{m_d^2-m_u^2}
\nonumber\eea
denotes a ratio of quark masses in pure QCD. One attitude is to use 
the measured decay width $\Gamma_{\eta\rightarrow 3\pi}$ for a determination 
of the quantity $Q^2$. On the
other hand, one may as well evaluate $Q^2$ from the meson mass ratio 
\bea
Q^2\! =\! \frac{m_K^2}{m_\pi^2}\frac{m_K^2-m_\pi^2}
{(m_{K^0}^2-m_{K^+}^2)_{\rm {QCD}}}(1+O(m_{quark}^2)),\!\!\!
\nonumber\eea
and predict the width. In 
this manner, the mass difference of the kaons in pure QCD shows up. 
In order to determine this difference, one has
to properly subtract the contributions from electromagnetic interactions
 to the kaon masses~\cite{dashen0}. Here one encounters a problem: due to
ultraviolet divergences generated by photon loops, the
splitting of the Hamiltonian of QCD+$\gamma$ into a strong and an
electromagnetic piece is ambiguous. The calculation of 
$(M_{K^+}^2-M_{K^0}^2)_{\rm QCD}$ in the effective theory must therefore 
reflect this ambiguity as well.
An analogous problem occurs whenever one wants to 
extract hadronic quantities from matrix elements which are 
contaminated with  electromagnetic contributions.

One is confronted  with two separate issues here. The first one 
is a proper definition of 
strong and electromagnetic contributions in a given theory. The second,
 separate point concerns the construction of the corresponding  
 effective low-energy Lagrangian (see also Ref. \cite{bijnens}).

The  aim  of this discussion is  i) to investigate the problem of
electromagnetic corrections in  QCD+$\gamma$, in the sense that 
the generating functional of Green functions of scalar, vector and axial
vector currents is extended to include radiative corrections at 
order $\alpha$, 
and ii) to construct the relevant effective theory at 
low energies, taking into account the ambiguities mentioned.
The Lagrangian built by Urech  \cite{Urech} so is worth a deeper study.

The problem is a complex one and I refer to the recent work \cite{GRS:03}  for a complete
discussion of some relevant examples and technical details.
In this work I will present an overview of ref.~\cite{GRS:03}.
\section{Parametrisation of the splitting}
\subsection{Some notation and tree level results}
In order to see how the splitting of strong and electromagnetic contributions works let me consider  as an example the linear sigma model (L$\sigma$M).
Without electromagnetism the Lagrangian of the model has an $O(4)$ symmetry spontaneously broken to $O(3)$.
The corresponding effective theory at low energies 
may  be analysed with the 
Lagrangian used in  ChPT, with low-energy constants  
 that are fixed in terms of the couplings 
of the L$\sigma$M \cite{GL,NS}.
Thus in this example the  L$\sigma$M acts as the strong high energy part of the theory.
I couple the four real scalar fields $\phi^A$ in the L$\sigma$M to  external 
vector and axial vector fields and incorporate  
electromagnetic interactions,
\eq\label{lagrV}
{\mathcal L}_{\sigma} \!\!\!&=&\!\!\!{\mathcal L}_0 
+{\mathcal L}_{\rm ct}\co
\nonumber\\
{\mathcal L}_0 \!\!\! &=& \!\!\!\frac{1}{2}\,(d_\mu\phi)^Td^\mu\phi
+\frac{m^2}{2}\,\phi^T\phi-\frac{g}{4}(\phi^T\phi)^2+c\phi^0
\nonumber\\&&
+\frac{\delta m^2}{2}\,(Q\cdot\phi)^T(Q\cdot\phi)\nonumber\\&&
-\frac{\delta g}{2}\,(\phi^T\phi)(Q\cdot\phi)^T(Q\cdot\phi)
\nonumber\\[2mm]
&&
-\frac{1}{4}\, F_{\mu\nu}F^{\mu\nu}-\frac{1}{2\xi}\,(\partial_\mu A^\mu)^2
\, ,
\en
The details of the notation and definitions can be found in ref.~\cite{GRS:03}.
What is important to note here is that
in our metric the spontaneously broken
 phase occurs at $m^2>0$. Since the electromagnetic 
interactions break 
 isospin symmetry, 
 we have explicitly introduced  the isospin breaking terms 
 $\sim \delta m^2, \delta g$ from the very beginning.  
 The counterterms are collected in 
 ${\mathcal L}_{ct}$, see ref.~\cite{GRS:03}.
The symmetry breaking 
parameter $c$ is considered
 to be of non-electromagnetic origin -- it provides the Goldstone bosons with a
 mass also at $e=0$. 
In order to render the  formulae more compact and make the counting more evident, I will use also 
the following notation for the
couplings $\dm$ and $\dg$,
\eq\label{e2}
\delta g=e^2gc_g\, ,\quad
\delta m^2=e^2m^2c_m\, ,
\en
where the new couplings $c_g$ and $c_m$ are assumed to be
independent of $e$ at this order,  $c_{g,m}\simeq O(p^0)$ and   $e^2,c \simeq
O(p^2)$.
At tree level the masses of the pions,  the sigma and the  vacuum expectation value are \eq\label{tree}
m_{\pi^0}^2 \!\!\!&=& \!\!\!\frac{c}{v_0}\co\;
m_{\pi^+}^2=m_{\pi^0}^2 - \dm +\dg v_0^2\co\nonumber\\[2mm]
m_\sigma^2\!\!\!&=& \!\!\! 2m^2+3m_{\pi^0}^2 \co\;\nonumber\\[2mm]
v_0\!\!\!&=& \!\!\!\frac{m}{\sqrt{g}}+\frac{c}{2m^2}+O(p^4)\fs
\en
  I omit here the issue of the splitting in the vector
currents for simplicity. Among the others the discussion of this issue is important
in order to understand the gauge dependence of the splitting in the
effective theory. I refer to \cite{GRS:03} for  this discussion.
\subsection{The splitting procedure and the matching scale $\mu_1$}
In order to illustrate the splitting procedure I will consider the
effect of the splitting  only on some  (running, physical) masses
of the model and also some strong coupling as $g$ and $c$. Strong effects are computed at one  loop and all the effects of
order $e^4$ are neglected.
For any kind of mass X (and also the strong couplings $g$ and $c$) it is possible to write
\eq
\label{eq:x0x1}
X=\bar X+e^2 X_1 \; ,
\en
where $\bar X$ is the pure strong part of the mass.
We want to define the pure strong contribution as that which is obtainable in a theory with $e=0$.
This definition is consistent at one loop with eq.~\ref{eq:x0x1} if
\eq\label{eq:xrun}
\left. \frac{d}{d\mu}X\right|_{e=0}= \frac{d}{d\mu}\bar X \; .
\en
This equation defines the dependence of $\bar X$ on the renormalization scale
$\mu$. This relation shows  also that one has to fix a boundary condition in order
 to fix $\bar X$. A   natural condition  consists in  choosing that  at a scale
$\mu_1$
\eq
\left. \bar X (\mu;\mu_1)\right|_{\mu=\mu_1}\equiv X(\mu_1)\fs
\en
The pure e.m. contribution comes then by the difference $X_1(\mu;\mu_1)=X(\mu)-\bar X (\mu;\mu_1)$.

To make things more explicit let us see the couplings $m$ and $g$ of the Lagrangian in eq.~\ref{lagrV}.
 The 
matching equations  are 
\eq\label{splt}
g(\mu)\!\!\!&=&\!\!\! \bar g(\mu ;\mu_1)
\biggl\{1+c_g\,\frac{e^2\bar g}{2\pi^2}\,
\ln\frac{\mu}{\mu_1}\biggr\}\, ,
\nonumber\\[2mm]
m^2(\mu)\!\!\!&=&\!\!\!\bar m^2(\mu ;\mu_1)\biggl\{1+(c_g+c_m)\,
\frac{e^2\bar g}{4\pi^2}\,
\ln\frac{\mu}{\mu_1}\biggr\}, \nonumber\\[2mm]
{c}\!\!\!&=&\!\!\!\bar c\fs
\en
In the following  I  denote with  a barred quantity an
expression evaluated at $e=0$, with $(g,m)\rightarrow (\bar g,\bar m)$.

\newcommand{\mpiz}{\mbox{${m_{\pi^0}^2}$}}
Another example is provided by the physical pion masses
($M_{\pi^{0,+}}$) at one loop.
To determine the physical pion masses, one evaluates the 
pole positions in the Fourier transform of the two-point
functions $\langle 0| T\phi^i(x)\phi^i(0)|0\rangle, i=1,3$.
In the following I will consider only the neutral pion mass for simplicity,
\eq\label{mpi}
M_{\pi^0}^2\!\!\!&=&\!\!\!m_{\pi^0}^2
\biggl\{1+\frac{g}{m_\sigma^2}\left(V_0+2L_{\pi^+}-L_{\pi^0}
\right)\biggr\}\nonumber\\[2mm] &&
+O(e^4,p^6)\fs
\en
where
\eq
V_0\!\!\!&=&\!\!\!(3+2y){L_\sigma}-\frac{m_\sigma^2}{48\pi^2}
(3+7y)\sem
y=\frac{m_{\pi^0}^2}{m_\sigma^2};\nonumber\\[2mm]
L_X\!\!\!&=&\!\!\!
\frac{m_X^2}{16\pi^2}\biggl\{\ln{\frac{m_X^2}{\mu^2}}
-1\biggr\}\fs \nonumber
\en
and $m_X$ are the tree level masses.
Starting from these equations one then expresses the parameters $g,m,c$  through the isospin
symmetric  couplings $\bar{g},\bar{m}$ and ${\bar c}$ by use 
of Eq.~(\ref{splt}). 
Next, we observe that the dependence on the electric charge in
 Eq.~(\ref{splt}) is an effect of order $\hbar$. 
 Therefore, to the accuracy considered here, 
the splitting (\ref{splt}) must be
applied to the tree-level expressions only,
\eq\label{dep_mu1}
v_0\!\!\!&=&\!\!\!\bar{v}_0\biggl\{1-C\ln{\mu^2/\mu_1^2}\biggr\}
+O(p^4)\co\nonumber\\[2mm]
\mpiz\!\!\!&=&\!\!\!
     \bar{m}_\pi^2\biggl\{1+C\ln{\mu^2/\mu_1^2}\biggr\}
+O(p^6)\scs
\en
where
\eq
C\!\!\!&=&\!\!\!(c_g-c_m)\frac{e^2\bar{g}}{16\pi^2}\scs
\bar{m}_\pi^2=\frac{\bar c}{\bar v_0}\fs
\en
 The $\mu_1$ dependence is
\eq\label{dep_mpiv0}
\mu_1\frac{d}{d\mu_1}(\bar{m}_\pi^2,\bar{v}_0)
=2C(\bar{m}_\pi^2,-\bar{v}_0)\fs
 \en

The strong part of $ M_{\pi^0}^2$ is the same for the neutral and for the 
charged pion mass,
\eq\label{eq:mpistrong}
\bar{M}_\pi^2&\doteq& \bar M_{\pi^0}^2=\bar M_{\pi^+}^2
\nonumber\\
\!\!\!&=&\!\!\!\bar{m}_\pi^2\biggl\{1+\frac{\bar{g}}{\bar{m}_\sigma^2}
(\bar{V}_0+\bar{L}_\pi)\biggr\}+O(p^6)\fs
\en
 The electromagnetic corrections 
 are  given by $e^2 M_{\pi^0}^{2,1}=M_{\pi^{0,+}}^2-
\bar{M}_\pi^2$. 
For the neutral pion mass they are 
\bea\!\!\!
e^2 M_{\pi^0}^{2,1}\!\!\!&=&\!\!\!
\frac{\bar m_\pi^2\bar{g}}{16\pi^2\bar{m}^2}
\biggl(
m_{\pi^+}^2\ln{\frac{m_{\pi^+}^2}{\mu^2}}-m_{\pi^0}^2
\ln{\frac{m_{\pi^0}^2}{\mu^2}}\biggr)\nonumber\\[2mm]\!\!\!
&&\!\!\! +\bar m_\pi^2 C\left(\ln{\frac{\mu^2}{\mu_1^2}}-1\right)
+O(e^4,p^6).
\eea
A similar expression holds for the charged pion 
mass.

The quantity $\bar{M}_\pi$ denotes the isospin symmetric 
part of the pion mass. It coincides neither with the neutral nor with the
charged pion mass, and is
independent of the running scale $\mu$. It depends, however,
 on the scale
$\mu_1$ where the matching has been performed,
\eq
\mu_1\frac{d}{d\mu_1}\bar{M}_\pi^2=2C\bar{m}_\pi^2
+O(e^4,p^6)\fs
\en
As $C$ is of order $e^2$, this scale dependence of the isospin
symmetric part is of  order $p^4$. The electromagnetic 
part $e^2M_{\pi^0}^{2,1}$ has the same scale dependence, up 
to a sign, as a result of which the total mass 
is independent of $\mu_1$.
\section{Splitting in the effective theory}
At low energy the L$\sigma$M  with the inclusion of electromagnetism can be
analyzed with the low energy effective theory of Gasser and Leutwyler, \cite{GL},
enlarged by Urech, \cite{Urech}.
The effective Lagrangian
 ${\cal L}_{\rm eff}$ is constructed from the Goldstone
 boson fields, the photon field and the external so\-ur\-ces
$r_\mu,l_\mu,f,$ and the spurion charges $Q_R$ and $Q_L$. The matching condition 
 states that the Green
 functions in
 the effective theory  must coincide with those in the
 original theory at  momenta 
much smaller than the $\sigma$-mass. 
At the end, one evaluates 
Green functions in the limit where the charge matrices 
become space-time independent.
Because the linear sigma model with space-time dependent 
spurion fields has
 the same symmetry as the theory that underlies the 
construction 
of the effective Lagrangian
 performed by Urech \cite{Urech}, 
by Mei\ss ner, M\"uller and Steininger \cite{MMS},  and by 
Knecht and Urech \cite{KU}, I  simply take over
their result. 
I will  determine  particular light energy constants (LECs)
by comparing physical quantities calculated
 in the underlying and in the effective theory. 
\subsection{Matching pion masses}
I first consider the purely strong part in the pion mass,
 displayed 
in Eq.~(\ref{eq:mpistrong}). For the low-energy expansion  one finds that 
\bea\label{eq:expstrong}
\bar M^2_\pi \!\!\!&=&\!\!\!\bar M^2\biggl[1-\frac{1}{32\pi^2}
\frac{\bar M^2}
{\bar F^2}
\left(     \frac{16\pi^2}{\bar g}
-11\ln\frac{2\bar m^2}{\mu^2} \right.
\nonumber\\[2mm]
&&+ \frac{22}{3} -\left.\ln{\frac{\bar M^2}{\mu^2}}   \right)
\biggr]+O(p^6)\scs
\eea
where the complete expression for $\bar F^2$ and $\bar M^2$ are reported in~\cite{GRS:03}.
 The quantity $\bar F$ denotes the pion decay constant in the
 chiral limit,
evaluated in the framework of the linear sigma model 
 at order $\hbar$, see Refs. \cite{GL,NS}, from where the 
 expression for $\bar F$ is taken.
 I have used the fact that $\bar M^2$ is 
linear in $c$ \cite{GL,NS} -- this fixes the structure 
of the expansion uniquely.

I may now compare Eq.~(\ref{eq:expstrong}) with the expansion 
 of the pion mass in the effective theory at $e=0$.
 I find for the 
parameters in the effective theory 
\bea\label{eq:matchBl3}
M^2\!\!\!&=&\!\!\!2\hat{m}B=\bar M^2\, ,\quad F^2=\bar F^2\co
\nonumber\\
l_3^r(\mu_{\rm eff})\!\!\!&=&\!\!\!-\frac{1}{64\pi^2}
\biggl( \frac{16\pi^2}{\bar g}
-11\ln\frac{2\bar m^2}{\mu^2} + \frac{22}{3}  \nonumber\\\!\!\!&&\!\!\!
+\ln\frac{\mu^2}{\mu_{\rm eff}^2}\biggr)\, ,
\nonumber\\[2mm]
l_7\!\!\!&=&\!\!\!0\, .
\eea
Note that $M^2, l_7$ and $F^2$ are independent of the scales
 $\mu$ 
and $\mueff$ of the underlying and of the effective
 theory.
On the other hand, the pion decay constant and the mass
 parameter $M^2$ depend
on the matching scale $\mu_1$. At one loop, 
\eq\label{mu1MF}
\frac{\mu_1}{F^2}\frac{d}{d\mu_1}\, F^2
\!\!\!&=&\!\!\!-2\frac{\mu_1}{M^2}\frac{d}{d\mu_1}\, M^2\nonumber \\
\!\!\!&=&\!\!\! 
\frac{e^2\bar g(c_m-c_g)}{4\pi^2}\fs
\en
The last term in this equation is proportional to the 
 char\-ged pion (mass)$^2$ in the chiral limit, see
 below. Using the DGMLY sum rule \cite{Dasetal} gives
\bea
\label{eq:fexp}
F(\mu_1=1\,\mbox{\small{GeV}} ) 
\!\!\!&=&\!\!\! F(\mu_1=500\,\mbox{\small{MeV}}) -0.1\, {\mbox{MeV}}.
\nonumber \\
\eea
The uncertainty related to $\mu_1$ so is of the order of the  PDG
 error \cite{Hagiwara}.

One can also determine the linear combinations
 ${\cal K}_{\pi^0}^r,$ 
of the electromagnetic couplings 
$k_i^r$ that occur in the expansion  of the neutral pion mass in the 
effective theory, see~\cite{GRS:03}. 
 One finds also that whereas the coupling
 ${\cal K}_{\pi^0}^r,$
 is independent of the scale $\mu$, 
 it depends on the matching scale $\mu_1$. 
Finally, I display the neutral pion mass in the linear
 sigma model,
properly expanded in powers of momenta, and electromagnetic
 corrections disentangled,
\bea
M_{\pi^0}^2\!\!\!&=&\!\!\!\bar M_\pi^2+e^2M_{\pi^0}^{2,1}+O(e^4)
\scs
\nonumber\\
\bar M_\pi^2\!\!\!&=&\!\!\!M^2\biggl\{1+\frac{2M^2}{F^2}\biggl(\,
l_3^r+
\frac{1}{64}\,\ln\frac{M^2}{\mu_{\rm eff}^2}\biggr)
\biggr\}\nonumber\\[2mm]\!\!\!&&\!\!\!
+O(p^6)\, ,\quad
\nonumber\\[2mm]\!\!
e^2 M_{\pi^0}^{2,1}\!\!\!&=&\!\!\!
    \frac{M^2}{16\pi^2F^2}
\left\{{ M_{\pi^+}^2}
\ln\frac{M_{\pi^+}^2}{\mu^2_{\rm eff}}-
M^2\ln\frac{M^2}{\mu^2_{\rm eff}}\right\}
\nonumber\\[2mm]\!\!\!
&&\!\!\!+e^2M^2{\cal K}_{\pi^0}^r+O(p^6)\fs
\eea

\subsection{A comparison  to other approaches within the model}
The splitting of pure e.m. and strong effects has been considered also
in other papers, see f.i. \cite{BP,Moussallam}.
What we want to discuss here is  the  approach presented here versus
the ones used previously.
To this aim
 I write 
the result (\ref{mpi}) for the neutral pion mass in the form
\bea
M_{\pi^0}^2&=&f_0+e^2f_1 +O(e^4,p^6)\co\nonumber\\
f_0&=&
m_{\pi^0}^2\biggl\{1+\frac{g}{m_\sigma^2}\,
(V_0+L_{\pi^0})\biggr\}\co
\nonumber\\[2mm]
e^2f_1&=&
2m_{\pi^0}^2\frac{g}{m_\sigma^2}\,\biggl\{L_{\pi^+}-L_{\pi^0}\biggr\}\fs
\eea
Since the physical mass is scale independent, one has
\bea\label{RG-M}
\mu\frac{df_0}{d\mu}=-e^2\mu\frac{df_1}{d\mu}\fs
\eea
Consider now the splitting of electromagnetic and strong
effects. In the language of  Ref. \cite{BP,Moussallam}, $f_0$ 
($e^2f_1$) is 
 the {\em strong} 
({\em electromagnetic}) part of the physical mass. 
 Both, the strong and the
electromagnetic parts of the mass   are $\mu$-dependent in 
this case.
One may again work out the low-energy representation 
of $M_{\pi^0}^2$ 
and identify the low-energy constants in this language. For
 the strong
part, one finds the expressions 
displayed in Eqs.~(\ref{eq:expstrong})-(\ref{eq:matchBl3}),
with
$(\bar g,{\bar m}^2) \to (g,m^2)$, whereas the 
electromagnetic LECs are collected in 
\bea\label{eq:k_imouss}
&&{\cal K}_{\pi^0}^r=
\frac{(c_g-c_m)g}{16\pi^2}\,\biggl(
\ln{\frac{\mueff^2}{\mu^2}}-1\biggr)\fs
\eea
Here, the $\mu$ dependence of ${\cal{K}}_{\pi^0}^r$ shows up.
This scale 
dependence of the electromagnetic part is canceled by 
the corresponding sca\-le dependence of the strong part.

In our framework, the {\em strong} part is given by
\bea
\bar M_{\pi}^2=f_0\biggr|_{g=\bar g,m=\bar m,c=\bar c}\co
\eea
where the couplings $\bar g,\bar m$ run with the 
strong part alone, see the discussion in earlier sections.
The difference $M_{\pi^0}^2-\bar M_{\pi}^2$ is called {\em
  electromagnetic correction} in this article. Both, 
the strong and the electromagnetic part are $\mu$-in\-de\-pen\-dent. 

We note that the $\mu$ dependence of 
${\cal K}_{\pi ^0}^r$  in
 Eq.~(\ref{eq:k_imouss}) is the same as the $\mu_1$ dependence 
in our procedure for the splitting. One can show that such a correspondence 
exists for all quantities that are $\mu$-independent. 
On the other hand, it does not hold anymore e.g. in the case of the
charged form factor, whose matrix elements are 
$\mu$ dependent. 

\section{Conclusions}
In this work  I have summarized some  of the main results
 of ref.~\cite{GRS:03}. 
In this work it is  outlined a method to  split consistently e.m. effects in 
Quantum Field Theory. 
The splitting  that is proposed is done order by order in the loop expansion. 
The strong part of a quantity depends only on couplings defined in a theory with $e=0$ 
(up to the desired perturbative order in $e$) and it has 
no running  proportional to the electromagnetic coupling $e$
(still, up to the perturbative order in $e$ which is considered).

In order to proceed correctly  in the construction of an effective theory it is important to
characterize the relevant  scales of the problem: 
$\mu$ (the renormalization scale of the underlying theory), $\mu_{\rm eff}$ (the renormalization scale of the effective theory) and $\mu_1$ (the scale at which the  strong part of a quantity is defined). 
The splitting ambiguities are parametrised by the scale
  $\mu_1$.  The uncertainty related  to $\mu_1$  can be of numerical
 relevance as it is shown in eq.~\ref{eq:fexp}.
In fact  the error induced  on $F$ by $\mu_1$ is of the order of the
 PDG error
\cite{Hagiwara}.

Another advantage of the splitting which is proposed here, is that
in the effective Lagrangian  the parameters in the strong sector
  are expressed through the ones of the underlying theory in its
  strong sector. This makes the matching between the underlying and
  the effective theory more transparent.

 Finally  the LECs of the effective  theory also contain all information about scale and
  gauge dependence of the Green functions in the underlying theory with electromagnetism.
\subsection{Acknowledgments}
I thank J. Gasser and A. Rusetsky for useful discussion.
This work has been partially supported by EC-Contract
HPRN-CT2002-00311 (EURIDICE).

\end{document}